\documentclass[runningheads]{llncs}

\usepackage{amsfonts}
\usepackage{amsmath}
\usepackage[T1]{fontenc}
\usepackage{graphicx}
\begin{document}
\title{A Serendipitous Recommendation System Considering User Curiosity}
%
%
\author{Zhelin Xu\inst{1,3} \and
Atsushi Matsumura\inst{2,3}}
\authorrunning{Z. Xu et al.}
%
\institute{Academic Service Office for the Library, Information and Media Sciences Area  \\ \email{asxzl1003@gmail.com} \\ \and
Faculty of Library, Information and Media Science \\ \email{matsumur@slis.tsukuba.ac.jp}\\ \and
University of Tsukuba, Tsukuba, Ibraki, Japan \\
}
\maketitle              
\begin{abstract}
To address the problem of narrow recommendation ranges caused by an emphasis on prediction accuracy, serendipitous recommendations, which consider both usefulness and unexpectedness, have attracted attention. 
However, realizing serendipitous recommendations is challenging due to the varying proportions of usefulness and unexpectedness preferred by different users, which is influenced by their differing desires for knowledge.
In this paper, we propose a method to estimate the proportion of usefulness and unexpectedness that each user desires based on their curiosity, and make recommendations that match this preference. 
The proposed method estimates a user's curiosity by considering both their long-term and short-term interests. 
Offline experiments were conducted using the MovieLens-1M 
dataset to evaluate the effectiveness of the proposed method. 
The experimental results demonstrate that our method achieves the same level of performance as state-of-the-art method while successfully providing serendipitous recommendations.
\keywords{Serendipity \and Recommender system \and Diversive curiosity}
\end{abstract}
\section{Introduction}\label{sec:intro}
With the rapid expansion of the Internet, the amount of information available to users on the Web has increased dramatically. 
However, the time required for users to make informed decisions has also grown, 
this significantly diminishing the user experience\cite{ricci2021recommender}. 
In order to solve this problem, recommendation systems based on data mining techniques have emerged since the mid-1990s.
Since then, numerous recommendation systems have been developed, aiming to provide valuable information to users.
Previous research on recommendation systems has assumed that user satisfaction is primarily determined by the system's ability to accurately predict and align with user preferences.
Consequently, efforts have focused on increasing user satisfaction by meeting this criterion\cite{ziarani2021serendipity}.
Thus, most traditional recommendation systems focus on prediction accuracy\cite{alhijawi2022survey}. 
However, these systems tend to recommend items that closely match the user's profile (e.g. action history and preferences)\cite{silva2018exploring}. 
This has exacerbated the filter bubble\cite{lunardi2020metric}, 
making it difficult for users to discover items unrelated to their profiles\cite{maccatrozzo2012burst}.
As a result, user satisfaction is low.

In response to this situation, 
it's not enough to focus only on prediction accuracy when building a recommendation system. 
Other evaluation indices are needed to evaluate the system\cite{mcnee2006being}.
In recent years, besides the prediction accuracy, 
evaluation metrics such as serendipity have gained significant attention \cite{kotkov2023rethinking}.
Serendipity is an evaluation metric that consists of unexpectedness and usefulness\cite{peng2020chestnut}. 
Serendipitous recommendations can free individuals from environments dominated only by items that align with their preferences, 
thereby increasing encounters with previously undiscovered or unexpectedly new items.
As a result, 
the problems associated with focusing primarily on prediction accuracy can be mitigated, 
leading to increase user satisfaction\cite{fu2023deep}. 

Since serendipitous recommendation makes user both satisfied and surprised, some studies have proposed serendipitous recommendation methods based on re-ranking algorithms\cite{peng2020chestnut,xu2023serendipity}. 
Re-ranking algorithms aim to find serendipitous items by combining a serendipity-oriented score (e.g. the item's unexpectedness score or unpopularity score) with the item's usefulness score\cite{kotkov2016survey}.
Moreover, some previous studies assume that the importance of the usefulness or serendipity-oriented score for candidate items are different. 
For instance, Li et al.\cite{li2020purs} attempted to weight the unexpectedness score of an item without considering the importance of usefulness. 
On the other hand, Kotkov et al.\cite{kotkov2020does} attempted to assign weights to both the usefulness and unexpectedness scores for each item.
However, they used the same predetermined weights for all users.

Unlike these previous studies, we consider that the ratio of usefulness to unexpectedness that each user expects from recommendations varies. 
In other words, since the desire for knowledge varies among users, some individuals seek new and novel information, while others prefer recommendations that closely align with their existing interests.
Therefore, to improve serendipity, it is necessary to provide recommendations that match the user's needs by assigning weights to the usefulness and unexpectedness scores of each candidate item. 

The desire for knowledge is closely linked to users' curiosity\cite{wang2020impacts}. 
In particular, 
people with high-curiosity tend to seek novel information, 
while those with low-curiosity prefer content related to their existing preferences. 
Therefore, in this paper, we propose a serendipitous recommendation system that consider user's curiosity. 
Specifically, we first calculate the usefulness and the unexpectedness scores of each item separately. 
These scores are then weighted according to each user's degree of curiosity, aiming to better match their unique desire for knowledge. This approach ensures that recommendations are more personalized.
Moreover, although the Curiosity and Exploration Inventory-II (CEI-II) has been used to estimate user's curiosity\cite{kashdan2009curiosity}. 
However, it relies on questionnaire-based analysis and requires users to answer multiple questions.
Consequently, CEI-II cannot automatically estimate the curiosity of a large number of users. 
To address this concern, we utilize user's short-term preference and long-term preference to automatically estimate their curiosity. 
Finally, we conduct an offline experiment on a public dataset, and results demonstrate that our method successfully provides serendipitous recommendations to users. 

\section{Related Work}

\subsection{Serendipitous Recommendation}
Several methods have been proposed to overcome the prediction accuracy limitations of traditional recommendation systems. 
Among these methods, some studies have focused on recommending highly unexpected items for users to facilitate serendipitous encounters with new information\cite{li2020directional}.
However, since such items often differ significantly from the user’s tastes, 
they may fall outside the user’s realm of knowledge and be completely unfamiliar, resulting in a lack of interest from the user\cite{maccatrozzo2017sirup}.

Re-ranking algorithms are a popular approach for making serendipitous recommendations. 
As described in the previous section, these algorithms aim to find serendipitous items in a prediction accuracy-oriented list by re-ranking this list.
Specifically, the usefulness score for each item is first predicted using a recommendation algorithm focused on prediction accuracy. 
These scores are then used to create a ranked list. 
Next, a serendipity-oriented score is calculated for each item and added to the usefulness score to obtain the final serendipity score. 
The list is re-ranked based on these serendipity scores, 
and the top items are recommended to the user as serendipitous items.

In this paper, we propose a re-ranking algorithm-based method to calculate each item's serendipity score and further match the users' needs based on their curiosity. 
This method allows for the recommendation of items that significantly diverge from the user's preferences when their curiosity is high, and items that closely align with their preferences when their curiosity is low.

\subsection{Curiosity in Recommendation Systems}
Curiosity theory has been applied in serendipity approaches in recent years. 
Menk et al. proposed a serendipitous tourism recommendation system by utilizing human curiosity, education level, gender and other information extracted from users' Facebook profiles\cite{menk2017curumim}. 
Although their method adjusts recommendations to include different levels of unexpectedness based on human curiosity, it categorizes curiosity into only three broad categories: slight, moderate and extreme, rather than treating it as a continuous value. 
Maccatrozzo et al. proposed a serendipitous TV program recommendation model that attempts to recommend some novelty items that are acceptable based on the user's curiosity\cite{maccatrozzo2017sirup}. 
Niu et al. posited that serendipity increases if the user's curiosity is aroused\cite{niu2017framework}.
In addition, the unexpectedness of the recommended item will affect the generation of the user's curiosity.
Consequently, they proposed a framework consisting of three components: surprise, value, and curiosity. 

The methods described above both require user feedback to estimate user's curiosity. 
Therefore, it is challenging to estimate the curiosity of a large amount of users simultaneously. 
In order to address this concern, 
in this paper, we use users' implicit feedback (e.g., clicks and shares)\cite{kelly2003implicit} to infer long-term and short-term preferences. 
We then automatically estimate curiosity based on these preferences.

\section{Proposed Method}
This chapter describes a serendipitous recommendation method based on user curiosity.
Section 3.1 explores the concept of curiosity, which is the focal point of this study. 
Next, in Section 3.2, we describe a method for estimating users' degree of \textit{diversive curiosity}. 
Section 3.3 then outlines a method for calculating the predicted usefulness and unexpectedness scores for each candidate item. 
Finally, Section 3.4 describes a method for making recommendations based on the user's curiosity, usefulness, and unexpectedness scores of items.

\subsection{Curiosity}\label{subsec:curiosity}
It has been noted that a user's degree of curiosity significantly influences their enjoyment when encounter something new and expected\cite{wang2020impacts}.
For example, even if an unexpected item is recommended to multiple users, 
the level of enjoyment experienced will differ because each user's level of curiosity is different.
In other words, we can determine whether users require new knowledge or not based on their curiosity.

Curiosity is an individual trait in the field of psychology and can be broadly classified into two types: epistemic curiosity and perceptual curiosity\cite{berlyne1960conflict}.
Epistemic curiosity is the desire for knowledge that motivates individuals to learn new knowledge and fill in information gaps\cite{litman2008interest}.
Perceptual curiosity also involves the desire for knowledge, 
but these two types of curiosity differ in the types of stimuli that activate these emotional states\cite{collins2004measurement}.
Epistemic curiosity is stimulated by complex ideas or conceptual ambiguity (e.g., a puzzle), 
while perceptual curiosity is stimulated by complex or ambiguous sensory experiences (e.g., pictures or sounds).

\begin{figure}[htb]
    \centering
    \includegraphics[width=\linewidth]{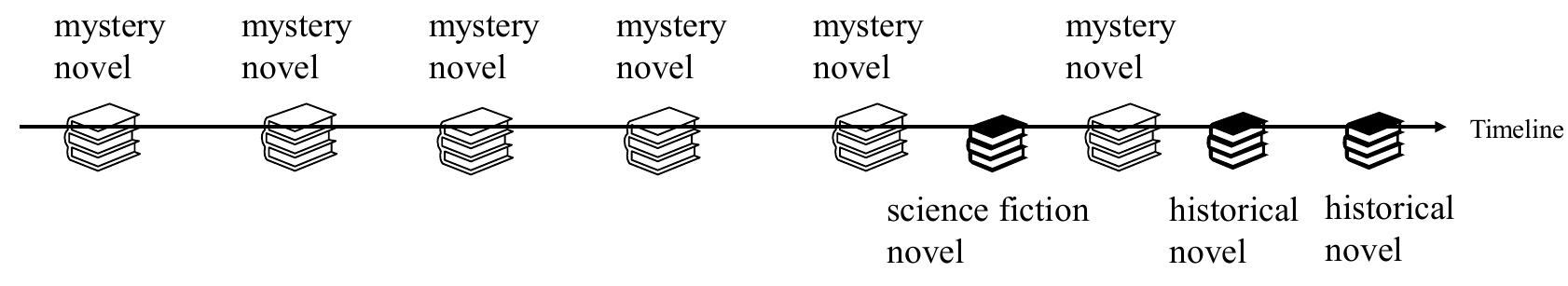}
    \caption{An example of behavior driven by diversive curiosity}
    \label{fig:diversive_curiosity}
\end{figure}

Epistemic curiosity can be further classified into two types: 
diversive curiosity and specific curiosity.
Diversive curiosity is a behavior that explores a wide range of new information and knowledge without a specific direction.
An example of diversive curiosity-driven behavior\cite{nishikawa2015} is shown in Figure \ref{fig:diversive_curiosity}, 
where a reader always reads mystery novels, but when bored by this genre, they seek out other types of novels such as science fiction or historical novels. 
On the other hand, specific curiosity involves directed exploration in response to contradictions.

As mentioned in Section \ref{sec:intro}, serendipitous recommendation aims to create unexpected encounters for users. 
Therefore, in this paper, we focus on the concept of diversive curiosity. 
After assessing each user's degree of diversive curiosity, we recommend new and unexpected items to those with higher levels of this trait. 
Conversely, for users with lower diversive curiosity, we recommend items that closely match their existing preferences and are less surprising. 
This approach ensures that the recommendations cater to each user's desire for knowledge.

\subsection{Modeling of Diversive Curiosity}\label{subsec:model_curiosity}
As mentioned in Section \ref{subsec:curiosity}, 
diversive curiosity refers to the behavior of exploring a wide range of new information and knowledge without a specific direction.
Based on this definition, 
the degree of user's diversive curiosity can be modeled by measuring the novelty and diversity of the information they seek. 
In this paper, we assess these characteristics by examining user's short-term and long-term preferences.
Short-term preferences refer to recent activities, 
while long-term preferences encompass what users have been engaged with over a longer period.

Drawing on the example of diversive curiosity-driven behavior\cite{nishikawa2015}, we determine the novelty of the information a user recently seeks by comparing the difference between their long-term and short-term preferences. 
Specifically, if there is a significant difference between a user's long-term and short-term preferences, 
such as a reader who usually reads mystery novels but has recently started reading science fiction novels or historical novels, we can infer that the user is seeking high-novelty stimuli to enhance their situation, as shown in Figure \ref{fig:diversive_curiosity}, 
Furthermore, we assess whether the user is seeking a broader range of information by measuring the diversity of their short-term preferences.
We focus on the diversity of short-term preferences rather than long-term ones, as our aim is to capture the user's current curiosity based on recent behavior.

Assuming that novelty and diversity of the information are equally important, diversive curiosity is hereby modeled as:
\begin{equation}
    \label{eq:curosity}
    Curosity_u = \frac{Diff_{u}^{long-short} +Div_{u}^{short}}{2} 
\end{equation}
Here, ${Diff_{u}^{long-short}}$ represents the difference between user $u$'s long-term and short-term preferences, 
which we use to determine the novelty of the information sought by the user. 
$Div_{u}^{short}$ indicates the diversity of user $u$'s short-term preferences.
In the following Section, we will detail the methods for \textit{calculating the difference between long-term and short-term preferences} and \textit{determining the diversity of short-term preferences}. 

\subsubsection{Calculating the Difference between Long and Short-term Preferences}
\begin{figure}[htb]
    \centering
    \includegraphics[width=\linewidth]{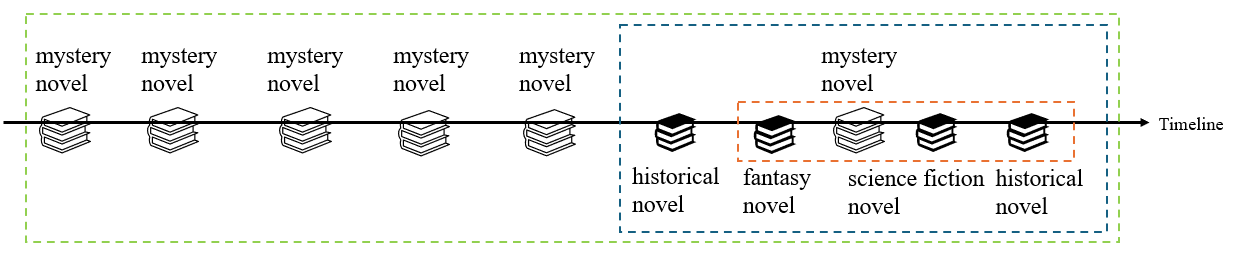}
    \caption{User sequence created from the user behavioral history. Left to right indicates the timeline from past to present.}
    \label{fig:sequence}
\end{figure}

The user's long-term and short-term preferences are modeled by considering temporal information. 
Initially, time-series data, as illustrated in Figure \ref{fig:sequence}, is generated from user's behavior sequence based on timestamps. 
Long-term preference is characterized by its consistency, as it reflects the activities that users have continuously engaged in over time\cite{yu2019adaptive}.
Therefore, we adopt a traditional method, singular value decomposition (SVD) to model the long-term preference from the user's entire behavioral history, which is indicated by the green box in Figure \ref{fig:sequence}). 
Behavioral history refers to the historical records of a user's activities, such as purchase history or movie rating records.

On the other hand, short-term preferences are defined as recent activities. User sequences are formed by their behavioral history.
It is suggested that short-term preferences can be modeled by analyzing the current $x$\% session of user sequences\cite{zhu2021neural}.
However, there is no consensus on what $x$\% should be.
For instance, as shown in Figure \ref{fig:sequence}, using either the five actions in the blue box or the four actions in the orange box are both viable methods,
each method yielding different results.
Therefore, in this study, we treat the proportion of behavioral history used to model short-term preferences as a parameter.

There are two primary methods for modeling short-term preference. 
Early work typically use Markov chains, while recently research has focused on recurrent neural networks (RNN)\cite{yu2019adaptive}. 
Compared to the Markov chain method, the RNN method accounts for the time sequence of the user's behavior, allowing it understand dynamic changes in preference\cite{zhu2021neural}. 
Therefore, we model the user's short-term preference using the Long Short Term Memory (LSTM) model.
LSTM is a type of RNN, has the following advantages in the improved model we adopted\cite{yu2019adaptive}:

\begin{enumerate}
    \item \textbf{Accounting for the time irregularities in the user's activities.} For example, if a user purchased product A on February 1, 2023, product B on February 2, 2023, and product C on May 10, 2024, it can be assumed that products A and B are closely related due to the close proximity of the purchase dates. 
    Therefore, the model includes a mechanism to assign higher relevance to products purchased within a short time frame.
    \item \textbf{Considering the semantics of the user's behavioral history. } For instance, if a user's purchase history includes items like an iPhone, AirPods, an umbrella, and a pencil case, these products each belong to different categories. When predicting the next purchase, particularly if it is an IT product, the model gives more weight to information related to the iPhone and AirPods, which belong to the same category.
\end{enumerate}

Using the two methods described above, we estimate 20 long-term and 20 short-term preferences for each user, creating sets of long-term and short-term preferences. 
When measuring the similarity between documents, the vector representing a document can be expressed as the average of the word vectors in the document.
Consequently, the similarity between documents can be measured using these vectors\cite{farouk2019measuring}. 
Following this approach, we calculate the vector representing the user's long-term interest set using Equation \ref{eq:interests_vec_long}.
\begin{equation}
    \label{eq:interests_vec_long}
        \text{Vector}_{u}^{\text{long}} = \frac{\sum_{i \in LP^{u}}\text{vector}_{u}^{i}}{20}
\end{equation}
\( LP^u \) represents the long-term preference set of user \( u \),
and \( vector_u^i \) represents the vector of the \( i \)-th preference in the user's long-term preference set. 
Similarly, we calculate the vector representing the user's short-term preference set using Equation \ref{eq:interests_vec_short}.
\begin{equation}
    \label{eq:interests_vec_short}
        \text{Vector}_{u}^{\text{short}} = \frac{\sum_{i^{\prime} \in SP^{u}} \text{vector}_{u}^{i^{\prime}}}{20}
\end{equation}
\( SP^u \) represents the short-term preference set of user \( u \),
and \(vector_u^{i^{\prime}}\) represents the vector of the \( i^{\prime} \)-th preference in the user's short-term preference set.
Each preference in long-term and short-term preference set is represented by an 80-dimensional vector. 
After normalizing the vector, the difference between the user's long-term and short-term preferences is computed by Euclidean distance:
\begin{equation}
    \label{eq:diff_long_short}
    Diff_{u}^{long-short}=\sqrt{\sum_{k}^{80}(Vector_{k}^{long}-Vector_{k}^{short})^2 }
\end{equation}
\( Vector_{k}^{long} \) or \( Vector_{k}^{short} \) is the \( k \)-th dimension of the vector representing user \( u \)'s long-term or short-term preference set.
\subsubsection{Determining the Diversity of Short-term Preferences}
We adopt the Intra-List Similarity\cite{zhang2012auralist} to compute the diversity of user's short-term preferences:
\begin{equation}
    \label{eq:div_short}
    Div_{u}^{short} = 1 - \overline{Intra-List\;Similarity}
\end{equation}
$\overline{Intra-List\;Similarity}$ is computed by: 
\begin{equation}
    \label{eq:intra-list}
    \overline{Intra-List\;Similarity} = \frac{1}{|S|} \sum_{u \in S} \sum_{\substack{m,n \in SP_{u,20}, \\ n < m}} CosSim(m, n)
\end{equation}
 \( S \) represents the total number of users. \( SP_u \) denotes the short-term preference set of user \( u \), which includes 20 preferences.
\( m \) and \( n \) are the individual short-term interests of user \( u \).
 \( CosSim(m, n) \) is the similarity between \( m \) and \( n \). 
 It is computed as follows:
 \begin{equation}
    \label{eq:cossim}
    CosSim(m, n) = \frac{|U_{m} \cap U_{n}|}{\sqrt{|U_{m}|} \times {\sqrt{|U_{n}|}}}
\end{equation}
\( |U_{m}| \) or \( |U_{n}| \) represent the number of users who have a short-term preference for $m$ or $n$, respectively. 
$|U_{m} \cap U_{n}|$ represents the number of users who have both short-term preferences $m$ and $n$. 
\subsection{Modeling of Usefulness and Unexpectedness}
The usefulness and unexpectedness of candidate items are modeled by adopting the method proposed in \cite{li2020purs}. 

\textbf{Modeling Usefulness.\quad}
The usefulness score of an item refers to the probability of matching the user's taste, which is calculated as:
\begin{equation}
    \label{eq:useful}
    useful_{u}^{i} = MLP(R_{u};E_{u};E_{i})
\end{equation}
where $E_{u}$ and $E_{i}$ are the embedding vectors representing user $u$ and the recommended item $i$, repectively. 
These vectors are obtained by the MLP network. 
$R_{u}$ is the embedding vector representing the user's behavioral history. 
The Click-Through-Rate model that combines a Bidirectional Gated Recurrent Unit (GRU) neural network and a self-attentive mechanism is trained for embedding the user's behavioral history. 

The Bidirectional GRU is a type of RNN that captures the temporal information and sequential order of the user's behavioral history. 
Compared to a traditional GRU, the Bidirectional GRU not only predicts future behaviors by learning from past actions but also learns future behavior to predict past actions.
This enables the model to consider both past and future information simultaneously when making predictions.
Additionally, the GRU is computationally more efficient than the LSTM due to reducing one gate. 
The self-attentive mechanism assigns weights to each action in the behavior of the user, which allows the model to capture the importance of both long-term and short-term preferences\cite{he2021locker}. 
After embedding these three elements, these embeddings are concatenated and put into an MLP network to predict the usefulness score.

As described in Section \ref{subsec:model_curiosity}, we chose to adopt an improved LSTM model to predict the user's short-term preferences, rather than using a model that combines a bidirectional GRU and a self-attentive mechanism, for the following reasons:
\begin{enumerate}
    \item LSTM outperformed GRU based on a real-world dataset\cite{yu2019adaptive}.
    \item Even with the use of a self-attentive mechanism, the user's short-term interests could not be modeled with high accuracy\cite{he2021locker}.
\end{enumerate}

\textbf{Modeling Unexpectedness.\quad}
The unexpectedness score of an item indicates the extent to which a user feels surprised by the recommended item. 
It can be accurately calculated based on the distance between the candidate item and the user's behavioral history in the feature space\cite{adamopoulos2014unexpectedness}.
The reason for this is that by embedding the recommended items and the user's behavioral history in a feature space, specifically a latent space, that can capture the complex relationships between them\cite{li2020latent}. 
The method described above for modeling the usefulness of items allows embedding in this latent space and transforming into vectors.
Using these vectors, the unexpectedness score of an item is calculated as:
\begin{equation}
    \label{eq:unexp}
    unexp_{u}^{i} = \sum_{k=1}^{N} d(w_i, C_k) \times \frac{|C_k|}{\sum_{k=1}^{N} |C_k|}
\end{equation}
Where $w_{i}$ is the vector of item $i$. 
$C_k$ represents the clusters obtained by applying the mean shift method to the user's behavioral history.
$|C_k|$ denotes the number of items within each cluster that are part of the user's behavior history. 
$d(w, C_k)$ represents the Euclidean distance between the recommended item $i$ and the cluster $|C_k|$ in the latent space.

\subsection{Modeling Serendipity}
As described in \ref{subsec:curiosity}, users with high degree of diversive curiosity tend to seek new information to expand their horizons.
Therefore, highly unexpected items that differ significantly from the user's interests will be recommended to them. 
On the other hand, users with low diversive curiosity are less likely to seek novel items, and thus, items that are only slightly surprising but still closely with their interests should be recommended. 
To achieve this goal, the user's diversive curiosity is used as a weight for both the usefulness and the unexpectedness of each item. 
Finally, the serendipity score of an item is calculated as:
\begin{equation}
    \label{eq:serendipity}
    serendipity_{u}^{i} = (1 - curiosity_{u}) \times useful_{u}^{i}
    + curiosity_{u} \times unexp_{u}^{i}
\end{equation}

\section{Experiments}
\subsection{Dataset}\label{subsec:dataset}
We utilized MovieLens-1M\cite{harper2015movielens} for our experimental analysis. 
This dataset includes movie titles, user ratings, and additional information. 
Today, it is widely recognized and used as a benchmark dataset in the field of recommender systems.
To obtain training, validation, and test datasets, 
we followed the leave-last-out data split method as described in \cite{he2021locker}. 
For example, if there are N items in a user's behavior sequence, 
the last one (($N$)th) is added to the test dataset, the ($N-1$)th item is added to the validation dataset, and the remaining ($N-2$) items are all added to the training data.
For each positive sample in the validation and test datasets, 
9 and 49 negative samples are randomly generated, respectively.

\subsection{Baseline Method}
In this paper, we used PURS\cite{li2020purs} as the baseline method. 
PURS is a state-of-the-art serendipitous recommendation model based on re-ranking algorithms. 
This model calculates the usefulness score and the unexpectedness score for each candidate item, respectively. 
Then assigns weights to the unexpectedness score based on the user's tolerance for unexpected items.
Our method models the diversive curiosity of each user and applies it as a weight to both the usefulness score and the unexpectedness score to calculate the serendipity score for each candidate item. 
Consequently, our approach differs from the PURS model.

\subsection{Evaluation Metrics}
As described in Section \ref{subsec:dataset}, we use MovieLens-1M to evaluate the performance of the proposed method.
Although this dataset contains user rating information, which can be used to assess users' tastes, it lacks data on serendipity. 
Therefore, to evaluate the serendipity performance of the proposed method, we indirectly measure the usefulness and unexpectedness of the recommended items generated by the model.
Specifically, precision and recall are used to evaluate the usefulness of the recommended items. 
Equation \ref{eq:unexp} is used to calculate the unexpectedness of the recommended items. 
Note that the unexpectedness scores of the candidate items in the test dataset range from 0.1024 to 0.2564. 

\subsection{Results}
Table \ref{tb:precision} and \ref{tb:recall} present the overall performance of two models in terms of precision and recall respectively. 
These results indicate that the proposed method slightly reduces the usefulness of recommendations. 
Conversely, Table \ref{tb:unexp} shows the overall performance of the two models in terms of the unexpectedness score. 
These findings demonstrate that the proposed method slightly outperforms PURS in terms of unexpectedness.
By assigning the user's diversive curiosity as a weight to both the usefulness and unexpectedness scores of the candidate items, the proposed method achieves similar performance to PURS. 
Therefore, the proposed method can effectively generate serendipitous recommendations.
\begin{table}[htb!]
\centering
\caption{Performance comparison in terms of precision. We use the last 30\% of user sequences to model short-term preferences.}
\label{tb:precision}
\begin{tabular}{p{2.8cm} p{2.2cm} p{2.2cm} p{2.2cm} p{2.2cm}}
\hline\hline
Model & precision@5 & precision@10 & precision@15 & precision@20 \\ \hline
Proposed method & 0.7500 & 0.7151 & 0.6886 & 0.6664  \\
PURS  & \textbf{0.7545} & \textbf{0.7185} & \textbf{0.6919} & \textbf{0.6683} \\ \hline
\end{tabular}
\end{table}
\begin{table}[htb!]
\centering
\caption{Performance comparison in terms of recall.}
\label{tb:recall}
\begin{tabular}{p{2.8cm} p{2.2cm} p{2.2cm} p{2.2cm} p{2.2cm}}
\hline\hline
Model & recall@5 & recall@10 & recall@15 & recall@20 \\ \hline
Proposed method  & 0.0762 & 0.1405 & 0.1970& 0.2463 \\
PURS & \textbf{0.0766} & \textbf{0.1409} & \textbf{0.1978} & \textbf{0.2465}   \\ \hline
\end{tabular}
\end{table}
\begin{table}[htb!]
\centering
\caption{Performance comparison in terms of unexpectedness score.}
\label{tb:unexp}
\begin{tabular}{p{2.8cm} p{2.2cm} p{2.2cm} p{2.2cm} p{2.2cm}}
\hline\hline
Model            & unexp@5         & unexp@10        & unexp@15        & unexp@20        \\ \hline
Proposed method & \textbf{0.1839} & \textbf{0.1820} & \textbf{0.1807} & \textbf{0.1799} \\ \hline
PURS             & 0.1815          & 0.1803          & 0.1793          & 0.1785 \\ \hline
\end{tabular}
\end{table}

\begin{figure}[htb!]
    \centering
  \scalebox{0.85}{  
    \includegraphics[width=\linewidth]{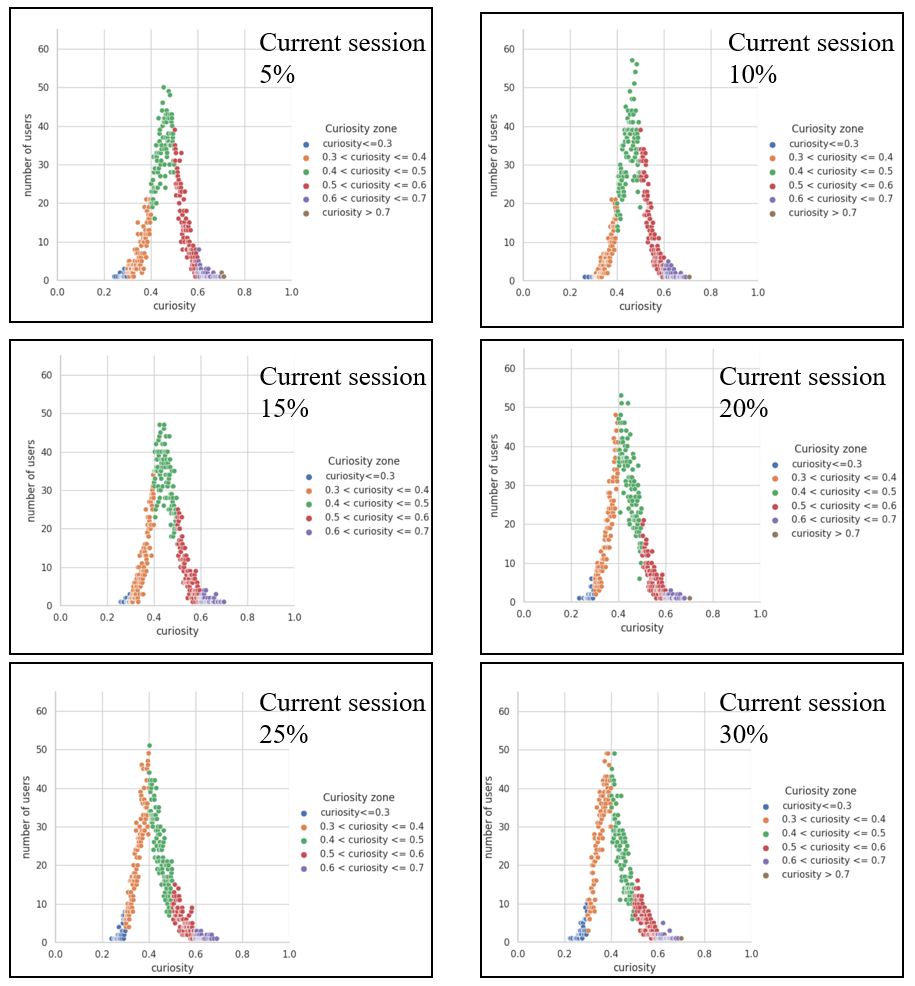}
    }
    \caption{Six scatterplots of users' diversive curiosity. ``5\%'' indicates that $x$ is set to 5.}
    \label{fig:curosity}
\end{figure}

Moreover, to enhance the performance of our proposed method, the approach for modeling short-term preferences needs to be improved. 
In this paper, we set $x$ as 30, which means we used the most recent 30\% of each user session (derived by historical behavior) for modeling. 
However, we hypothesize that changes in the value of $x$ impact the user's diversive curiosity since it may change the user's short-term performances. 
To validate this hypothesis, we also modeled user's diversive curiosity by setting $x$ into \{5, 10, 15, 20, 25\}.
The results are shown in Figure \ref{fig:curosity}. 
Nodes of different colors represent varying degrees of users' diversive curiosity; for example, a red node indicates a curiosity degree ranging from 0.5 to 0.6.  
From these results, we observe that as the $x$\% of the current session changed, the scatterplot of their diversive curiosity also changed. 
This means that some users' diversive curiosity changed as well.
Additionally, we found that the number of users with low diversive curiosity increased as the value of $x$ increased. 
Therefore, to provide more serendipitous recommendations, it is essential to develop a method for modeling diversive curiosity that uses the appropriate behavioral history for each user.

\subsection{A Case Study of Diversive Curiosity}
In this section, we present a case study of modeling a user's diversive curiosity on the MovieLens-1M using our proposed method.
This user's diversive curiosity was determined to be 0.7014, indicating a high degree of diversive curiosity. 
As illustrated in Section \ref{subsec:model_curiosity}, 
the degree of a user's diversive curiosity is determined by considering their long-term and short-term preferences.
Therefore, we will examine this user's long-term and short-term preferences to confirm whether they truly has diversive curiosity.
Table \ref{tb:long-term preferences} and \ref{tb:short-term preferences} show the long-term and short-term preferences of this user, respectively. 
From these results, we can observe the following two points:
\begin{enumerate}
    \item \textbf{The user's long-term preferences are different from short-term preferences.} 
    The user's long-term preference is to see comedy movies, which make up 45\% ($\frac{9}{20}$) of all movies. 
    In contrast, among the user's short-term preferences, comedy movies dropped to 25\% ($\frac{5}{20}$). 
    Additionally, the $westerns$ and $adventure$ movies that the user has recently watched are not part of their long-term preferences.
    \item \textbf{Diversity in short-term preferences.} 
    In the user's short-term preferences, they watched nine different types of movies. 
    Moreover, there is no significant number of movies watched in a specific genre.
\end{enumerate}

As described in Section \ref{subsec:curiosity}, diversive curiosity is a behavior that explores a wide range of new information and knowledge without a specific direction. 
Based on the observations above, we can conclude that the user's behavior in this case fits this definition. 
Moreover, the differences between long-term and short-term preferences, as well as the diversity within the short-term preferences, indicate that the user exhibits a broad range of interests and a willingness to explore different genres. 
This implies that such users prefer items containing elements of unexpectedness. By employing the proposed method, which incorporates curiosity as a weighting factor, we can recommend items that more effectively meet the needs of these users.

\begin{table}[htb!]
\centering
\caption{The user's long-term preferences list. Items at the top of the list are most likely to reflect the user's long-term preferences.}
\label{tb:long-term preferences}
\scalebox{0.7}{ 
\begin{tabular}{p{2cm}p{6.5cm}p{2.2cm}p{2cm}}
\hline\hline
Number & Movie Title       & Genre  & Release Date    \\ \hline
1 &Swingers  & Comedy & 1996  \\ \hline
2 &Big Sleep    & Film-Noir  & 1946     \\ \hline
3 &Glengarry Glen Ross  & Crime & 1992      \\ \hline
4 &Crimes and Misdemeanors & Comedy & 1989 \\ \hline
5 &Harvey & Comedy & 1950\\ \hline
6 &Stop Making Sense & Documentary  & 1984\\ \hline
7 &Brother from Another Planet, The &  Comedy & 1984 \\ \hline
8 &Chinatown & Film-Noir & 1974 \\ \hline
9 &Stars and Bars  & Comedy & 1988 \\ \hline
10 &Ed Wood  & Comedy & 1994  \\ \hline
11 &Key Largo  & Crime & 1948  \\ \hline
12 &Hustler, The  & Sport & 1961   \\ \hline
13 &Living in Oblivion  & Comedy & 1995   \\ \hline
14 &Grapes of Wrath, The  & Drama & 1940   \\ \hline
15 &City Lights & Comedy & 1931  \\ \hline
16 &Hard-Boiled (Lashou shentan)  & Action & 1992  \\ \hline
17 &Brazil & Sci-Fi & 1985   \\ \hline
18 &One Flew Over the Cuckoo's Nest  & Drama  & 1975  \\ \hline
19 &Hard Day's Night, A  & Comedy & 1964  \\ \hline
20 &Magnolia& Drama & 1999\\ \hline
\end{tabular}
}
\end{table}

\begin{table}[htb!]
\centering
\caption{The user's short-term preferences list.}
\label{tb:short-term preferences}
\scalebox{0.7}{ 
\begin{tabular}{p{2cm}p{6.5cm}p{2.2cm}p{2cm}}
\hline\hline
Number & Movie Title       & Genre  & Release Date    \\ \hline
1 &Far Off Place, A  & Adventure & 1993  \\ \hline
2 &Santa Fe Trail    & Western  & 1940     \\ \hline
3 &Inventing the Abbotts  & Romance & 1997      \\ \hline
4 &Omega Man, The  & Action & 1971 \\ \hline
5 &Urbania & Drama & 2000\\ \hline
6 &Tombstone & Western  & 1993\\ \hline
7 &Lost \& Found &  Romance & 1999 \\ \hline
8 &Best Men & Action & 1997 \\ \hline
9 &Bushwhacked & Adventure & 1995 \\ \hline
10 &Son of the Sheik, The  & Adventure & 1926  \\ \hline
11 &Crime and Punishment in Suburbia & Thriller & 2000  \\ \hline
12 &Journey of August King, The & Drama & 1995   \\ \hline
13 &Me Myself I & Comedy & 2000   \\ \hline
14 &Faculty, The & Horror & 1998   \\ \hline
15 &Cleo From 5 to 7 & Drama & 1962  \\ \hline
16 &Theodore Rex & Comedy & 1995  \\ \hline
17 &Reality Bites & Comedy & 1994   \\ \hline
18 &Idiots, The (Idioterne) & Comedy  & 1998  \\ \hline
19 &Mystery Science Theater 3000: The Movie & Comedy & 1996  \\ \hline
20 &Planet of the Apes& Action & 1968\\ \hline
\end{tabular}
}
\vspace{-11.5pt}
\end{table}

\section{Conclusion}
Recently, serendipitous recommendations have garnered significant attention. 
However, previous studies do not consider the user's disire for knowledge. 
Since the desire for knowledge is closely linked to users' curiosity, 
we propose a serendipitous recommendation method that considers user's diversiver curiosity. 
Specifically, the usefulness score and the unexpectedness score are first calculated for each candidate item. 
After that, we determine the degree of diversive curiosity for each user based on their long-term and short-term preferences. 
Finally, these scores are then applied as weights to the usefulness and unexpectedness scores to obtain the serendipity score of each candidate item. 

An evaluation experiment was conducted on the public dataset MovieLens-1M. 
The results showed that the proposed method achieved the same level of performance as a state-of-the-art method, successfully enabling serendipitous recommendations. 
Future work will focus on developing a method for more accurately modeling the user's diversive curiosity and conducting comprehensive experiments to validate the effectiveness of our method.

\begin{credits}
\subsubsection{\discintname}
The authors have no competing interests to declare that are
relevant to the content of this article. 
\end{credits}

%
%
%
\bibliographystyle{splncs04}
\bibliography{refs}
%




\end{document}